\documentclass[12pt,preprint]{aastex}
\input psfig.sty

\def\mpc{\,h^{-1}{\rm {Mpc}}}
\def\rhor{\rho(\vec{r})}
\def\deltr{\delta(\vec{r})}
\def\deltk{\delta(\vec{k})}

\begin{document}

\title{The non-linear evolution of bispectrum from the scale-free N-body simulation}

\author{Y.H. Hou${^{1,2,4}}$, Y.P. Jing${^{1,4}}$, D.H. Zhao${^{1,4}}$, G. B\"orner${^{3,4}}$}
\altaffiltext{1}{Shanghai Astronomical Observatory, the Partner Group of
MPI f\"ur Astrophysik, Nandan Road 80,  Shanghai 200030, China}
\altaffiltext{2}{Graduate School of the Chinese Academy of Science, Yu Quan Road 19a, Beijing 100039, China}
\altaffiltext{3}{Max-Planck-Institut f\"ur Astrophysik,
Karl-Schwarzschild-Strasse 1, 85748 Garching, Germany}
\altaffiltext{4}{e-mail: hyh@center.shao.ac.cn, ypjing@center.shao.ac.cn,
dhzhao@center.shao.ac.cn, grb@mpa-garching.mpg.de}
\begin{abstract}
We have accurately measured the bispectrum for four scale-free models
of structure formation with the spectral index $n=1$, $0$, $-1$, and
$-2$. The measurement is based on a new method that can effectively
eliminate the alias and numerical artifacts, and reliably extend the
analysis into the strongly non-linear regime. The work makes use of a
set of state-of-the art N-body simulations that have significantly
increased the resolution range compared with the previous studies on
the subject. With these measured results, we demonstrate that the
measured bispectrum depends on the shape and size of $k$-triangle even
in the strongly nonlinear regime. It increases with wavenumber and
decreases with the spectral index.  These results are in contrast with
the hypothesis that the reduced bispectrum is a constant in the
strongly non-linear regime. We also show that the fitting formula of
Scoccimarro \& Frieman (1999) does not describe our simulation results
well (with a typical error about 40 percent).  In the end, we present
a new fitting formula for the reduced bispectrum that is valid for $-2
\leq n \leq 0$ with a typical error of $10$ percent only.

\end{abstract}




\section{Introduction}
Large-scale structures in the Universe are thought to arise from small
 primordial fluctuations through gravitational amplification.  It is
 known that gravitational clustering is a non-linear process. When the
 density fluctuations are sufficiently small, the evolution of the
 structures can be studied using perturbation theory (PT). With the
 growth of the fluctuation, even for an initially Gaussian
 fluctuation, nonlinear gravitational instability induces non-Gaussian
 signatures in the density field. In the weakly non-linear regime,
 leading order (tree-level) perturbation theory (Juszkiewicz, Bouchet
 \& Colombi 1993; Bernardeau 1994a; Bernardeau et al. 1994b; \L okas
 et al. 1995; Gazta\~{n}aga \& Baugh 1995; Baugh, Gazta\~{n}aga \&
 Efstathiou 1995; Bouchet et al.  1995) can describe the clustering
 properties successfully. As one approaches smaller scales, the loop
 corrections to the tree-level results are expected to become
 important (Scoccimarro \& Frieman 1996; Scoccimarro 1997). In the
 non-linear regime, numerical simulations must be applied to follow
 the development of the cosmic structures.
 
The n-point correlation functions have been widely used as a powerful
tool for quantifying the statistical properties of a density field
both in theoretical models and observational catalogs (Peebles
1980). For a Gaussian random field, the two-point correlation function
(2PCF) or its Fourier transform, the power spectrum $P(k)$ can
completely characterize its statistical properties, with all
higher-order (connected) correlation functions being zero. It requires
the higher order correlation functions to describe the statistical
properties of the non-Gaussian distribution resulting from
gravitational instability (Peebles 1980; Fry 1984; Bernardeau et
al. 2002 for an excellent review and references therein).

The bispectrum, the three-point correlation function (3PCF) in Fourier
 space, is the lowest order statistic that probes the shape of
 large-scale structures generated by the gravitational clustering
 (Peebles 1980).  Theoretical models of weakly non-linear 3PCF have
 been studied well in the literature based on PT. PT can describe
 properties of dark matter on large scales $\ge 10 \mpc$. It predicts
 that the 3PCF depends on the shape of the linear power spectrum and
 on the shape of the triangle configuration both in real space (Jing,
 B\"orner \& Valdarnini 1995; Jing \& B\"orner 1997; Frieman \&
 Gazta\~naga 1999; Barriga \& Gazta\~naga 2002) and in Fourier space
 (Fry 1984; Scoccimarro et al. 1998; Scoccimarro et al. 1999).  When
 the galaxy bias is considered, the bispectrum of galaxies contains
 information on the primordial fluctuation and on galaxy biasing (Fry
 1994; Fry \& Gazta\~naga 1993; Hivon et al. 1995; Mo, Jing, \& White
 1997; Matarrese et al. 1997; Verde et al. 2002). Measuring the galaxy
 bispectrum on large scales can help break the degeneracy between the
 linear bias and the matter parameter $\Omega_m$ usually present in
 the dynamical analysis of galaxy redshift surveys (Fry 1984; Hivon et
 al. 1995; Matarrese et al. 1997; Verde et al. 1998; Scoccimarro et
 al. 1998).  Several authors have started to measure the bias
 parameters from current large galaxy surveys (Frieman \& Gazta\~naga
 1999 and Gazta\~naga \& Freiman 1994 for the APM galaxies;
 Scoccimarro et al. 2001c for IRAS galaxies; Verde et al. 2002 and
 Jing \& B\"orner 2003 for the 2dFGRS galaxies; Kayo et al. 2004 for
 SDSS galaxies).

A quantitative modeling of the 3PCF or bispectrum in the nonlinear
regime is more challenging. There have been attempts to predict the
3PCF based on the so-called halo model (Ma \&
Fry 2000, Scoccimarro et al. 2001b, Takada \& Jain 2003a, Wang et
al. 2004). In their detailed modeling, Takada \& Jain (2003a) found
that the halo model prediction for the 3PCF agrees with the simulation
result of Jing \& B\"orner (1998) both in linear and strongly
non-linear regimes, but fails on intermediate non-linear scales ($\sim
1\mpc$). They also pointed out that the 3PCF at the intermediate
scales is very sensitive to the outer radial cut of halos, which could
be the reason for the failure of the halo model. On the other hand,
N-body simulations have been widely used to study the 3PCF or
bispectrum in the nonlinear regime (Davis et al. 1985; Efstathiou et
al. 1988). Based on extensive studies with N-body simulations, a
fitting formula for the bispectrum was proposed for the scale-free
models by Scoccimarro \& Frieman(1999, hereafter SF99), and then
extended for the cold dark matter (CDM) models by Scoccimarro \&
Couchman (2001).

The fitting formula of Scoccimarro \& Couchman (2001) was applied to
calculating the skewness of the convergence field in the weak
gravitational lensing survey (Van Waerbeke et al. 2001; Hamana et
al. 2002).  Nowadays weak lensing surveys have become detailed to start
measuring the skewness of the lensing shear field (Pen et al. 2003;
Zhang et al. 2003), and will soon become big enough to measure the
three-point correlation function of cosmic shears (Schneider et
al. 2003; Takada \& Jain 2003b). It is important to reliably predict the
skewness and the three-point correlation function of cosmic shears in
cosmological models, as the observational determinations of these
quantities are expected to yield a measure of the cosmological density
parameter (Bartelmann \& Schneider 2001 for an excellent review).
Therefore, the motivation is high to derive an accurate model for the
bispectrum from linear to non-linear scales, and present them in a
form useful for the weak lensing survey analysis.

In this paper, we present such a study to investigate how the
non-linear evolution of the bispectrum proceeds with a set of
scale-free simulations of $512^3$ particles.  The simulations have the
initial spectral index $n=1$, $0$, $-1$, or $-2$. We find that the
bispectrum depends not only on $n$ but also on the shape and size of
the triangle even in the strong non-linear regime. Our results show
that the formula of SF99 cannot accurately describe the properties of
the bispectrum in the non-linear regime. Comparing our measured
non-linear bispectrum with the weakly non-linear bispectrum obtained
from the second order PT (hereafter PT2), we have arrived at a new
formula for the bispectrum that is significantly more accurate than
that of SF99.

In this analysis, we have carefully examined possible effects on the
bispectrum measurement of the numerical artifacts, such as finite box
size, force softening, and particle discreteness. We have also closely
paid attention to the effect caused by the mass assignment using the
Fast Fourier Transform (FFT, i.e. the alias effect). In order to get
the true power spectrum and bispectrum from the simulation, we have
developed a procedure to correct for the numerical and alias effects.

This paper is organized as follows. In section 2, we give a brief
overview of the self-similar evolution of the density field. In
section 3, we describe the numerical simulations used. In section 4,
we introduce the power spectrum and the bispectrum, and outline the
possible numerical artifacts in the measurements of the power spectrum
and the bispectrum. In section 5, we describe our method to measure
the bispectrum. We present the measured bispectrum, and give our
fitting formula for the bispectrum in section 6. The last section
contains our summary.
\section{Self-similarity}

Gravitational clustering from initial conditions has provided one of
the classic problems in cosmology. To achieve a self-similar
evolution, according to Peebles (1980) and Efstathiou et al.(1988),
two conditions must be fulfilled: (1) The background cosmology should
not contain any characteristic scales, thus the universe must be an
Einstein-de Sitter one, and (2) The initial density field should have
no characteristic length scale, thus its initial power spectrum must
be a power law.

For a self-similar clustering pattern, its physical properties
remain the same during its evolution when the length scale $R$ is scaled by a
characteristic scale $R_0$. A simple choice of $R_0$ for a
self-similar clustering is the scale at which the fluctuations begin
to become non-linear, i.e., the variance of the linear density field
smoothed on this scale is unity,
\begin{equation}
\sigma^2(R_0,a)=1\,.
\label{characterR0}
\end{equation}
The variance is defined by 
\begin{equation}
\sigma^2(R,a)=\int \Delta^2_L(k,a)|W(kR)|^2{dk \over k}.
\label{sigmar0}
\end{equation}
where $W(x)$ is the Fourier transform of a window function(usually a
top hat or Gaussian), $\Delta^2(k)=d\sigma^2/d\ln
k=(V_\mu/(2\pi)^3)4\pi k^3 P(k)$ is the contribution to the fractional
density variance per unit $\ln k$ ($V_\mu$ is a normalization volume),
and $a(t)$ is the expansion scale factor. For the scale-free initial
power spectrum, $P_L \propto a^2k^n$, the characteristic scale
satisfies
\begin{equation}
R_0 \propto a(t)^{2 \over 3+n} .
\end{equation}
The characteristic wavenumber $k_0$ can be chosen to be $ R_0^{-1}$,
 so $k_0 \propto a(t)^{-2/(3+n)}$.  With the characteristic scale
 $R_0$, all statistical measures of the density field can be expressed
 as a similarity solution that is independent of time
\begin{eqnarray}
f(R,t)=g(R/R_0) \;   \; or  \; \;   f(k,t) = g(kR_0),
\end{eqnarray}
(Peebles 1980; Efstathiou et al. 1988; Colombi, Bouchet, \& Hernquist 1996;
Jain \& Bertschinger 1998). 

\section{The numerical simulation}
We study the scale-free models that assume an Einstein-de Sitter
universe (i.e. $\Omega_0=1$ and $\Lambda_0=0$), and a power-law
$P(k)\propto k^n$ with $n$ being $1$, $0$, $-1$, and $-2$ respectively
for the linear density power spectrum. For each model, we have one
simulation of $512^3$ particles that was produced by one of the
authors (YPJ). The current simulations are constructed in a similar
way as the scale-free simulation sample in Jing (1998), but have
higher force and mass resolutions.  They were generated with a
parallel-vectorized ${\rm P^3M}$ (i.e. Particle Particle Particle
Mesh) code (Jing \& Suto 2002) at the National Astronomical
Observatory of Japan.  The gravitational force is softened with the
${\rm S2}$ form of Hockney and Eastwood (1981) with the softening
parameter $\eta = 1\times 10^{-4}L$ ($L$ is the simulation box size),
so the force becomes Newtonian when the separation is larger than
$\eta$. The simulations are evolved for 2000 time steps with a total
of ten (n=$1$, $0$, $-2$) or eleven (n= $-1$) outputs at a constant
logarithmic interval ($\Delta \log a$) in the scale factor $a$. Table
1 summarizes the parameters that are relevant to the discussion in the
current work.

\section{Power and bispectrum}

\subsection{Basic theory }
Let $\rhor$ be the cosmic density field, with the mean density
$\bar\rho$. The density field can be represented by a dimensionless
field $\deltr$ (which is usually referred to as the density contrast)
\begin{equation}
\deltr=\frac{\rhor - \bar\rho}{\bar\rho}.
\label{rhor1}
\end{equation}
Based on the cosmological principle, we expect $\rhor$ to be periodic
in some large rectangular volume $V_\mu$. Its Fourier transformation is
then defined by
\begin{equation}
  \deltk = \frac{1}{V_\mu} \int_{V_\mu} \deltr e^{i\vec{r} \cdot \vec{k}}d\vec{r}.
\label{deltak1}
\end{equation}

The density field of the simulation is periodic at the box size
$L$. The requirement of periodicity restricts the allowed wavenumbers
to harmonic boundary conditions
\begin{equation}
k_x=nk_b  ,(k_b={2 \pi \over L} , n=\cdots-2,-1,0,1,2,3\cdots ),
\label{wavenumberx}
\end{equation}
with similar expressions for $k_y$ and $k_z$.

The power spectrum $P(k)$ and the bispectrum
$B_{123}=B(\vec{k}_1,\vec{k}_2,\vec{k}_3)$ are defined as
\begin{eqnarray}
\langle \delta(\vec{k}_1)\delta(\vec{k}_2) \rangle  &=& \delta_{Dirac}(\vec{k}_1+\vec{k}_2)P(k), \cr 
\langle \delta(\vec{k}_1)\delta(\vec{k}_2)\delta(\vec{k}_3) \rangle &=&  \delta_{Dirac}(\vec{k}_1+\vec{k}_2+\vec{k}_3)B_{123} ,
\label{powbisp}
\end{eqnarray}
where $\langle \cdots \rangle$ means ensemble average, $\delta_{Dirac}$
is the Dirac delta, and the
$\delta_{Dirac}(\vec{k}_1+\vec{k}_2+\vec{k}_3)$ implies that the
bispectrum is defined for configurations of wavenumbers that form
closed triangles in $k$-space.  There are many ways to express the
shape of a triangle. For a triangle with $\vec{k}_1$, $\vec{k}_2$,
$\vec{k}_3$, and $|\vec{k}_1| \geq |\vec{k}_2|$, we can parameterize
its shape by $k$, $v$, $\theta$ as:
\begin{eqnarray}
 k=|\vec{k}_1|,\ \ \  v={|\vec{k}_1| \over |\vec{k}_2|}, \  \  \ \theta=\arccos \left( {\vec{k}_1 \cdot \vec{k}_2 \over |\vec{k}_1| |\vec{k}_2|} \right ).
\label{shapetriangle}
\end{eqnarray}
In the tree-level PT, the bispectrum can be expressed as follows:
\begin{equation}
B_{123}=2~F_2(\vec{k}_1,\vec{k}_2)P_1 P_2 +{\rm cyc.} ,
\label{B123}
\end{equation} 
where $P_i \equiv P(\vec{k}_i)$ ($i=1,2,3$) , and $F_2(\vec{k}_1,\vec{k}_2)$ is the kernel function,
\begin{equation}
F_2(\vec{k}_1,\vec{k}_2)={5\over 7}+{1\over 2}{\vec{k}_1 \cdot \vec{k}_2 \over
 k_1 k_2}\left( {k_1\over k_2}+{k_2\over k_1}\right)+{2\over 7}\left( 
{\vec{k}_1 \cdot \vec{k}_2 \over k_1 k_2}\right)^2  .
\label{F2kerl}
\end{equation}
For convenience we can define the reduced bispectrum $Q$ as 
\begin{equation}
Q(\vec{k}_1,\vec{k}_2,\vec{k}_3)={B_{123}\over P_1 P_2 +P_2 P_3 +P_3 P_1}.
\label{Q123}
\end{equation}
According to Eq. (\ref{shapetriangle}), $Q$ can be expressed as a
function of $k$, $v$, and $\theta$.

\subsection{Measuring the bispectrum}

The Fourier modes of a particle distribution can be determined
exactly using the expression (Peebles 1980)
\begin{equation}
\delta(\vec{k})={1 \over N}\sum_{i=1}^Ne^{i\vec{k}\cdot \vec{x}_i}.
\label{fourdeltadir}
\end{equation}
Owing to the periodic boundary condition in the simulation,
wavenumbers are restricted to the form defined by
Eq.~(\ref{wavenumberx}). It is inefficient to use
Eq.~(\ref{fourdeltadir}) to compute the Fourier transformation for a
simulation where both $N$ and the mode number considered are
large. Here we use the Fast Fourier Transform (FFT) technique to
compute $\deltk$. In this case, there is an upper limit for $\vec k$
that is imposed by the finite sampling of the density field at the FFT
mesh points, which is called the Nyquist wavenumber,
\begin{equation}
k_{Ny}={\pi \over \Delta x} ,
\end{equation}
where $\Delta x=L/N_m$ is the mesh spacing and $N_m$ is the dimension
of the mesh. Then the power spectrum and bispectrum can be estimated
through averaging all modes of which the wavenumbers $k_i$ in thin
shells ($k_i \pm \Delta k $, $\Delta k \ll k_i $,
i=1, 2, 3) satisfy Eq.~(\ref{powbisp}):
\begin{eqnarray}
 {\hat{P}(k_1)} &=&{1 \over m} \sum_{\vec{k}_1, \vec{k}_2 \in \Phi} 
(\delta(\vec{k}_1)\delta(\vec{k}_2)) \delta_{Dirac}(\vec{k}_1+\vec{k}_2),\nonumber\\
{\hat{B}_{123}} &=& {1 \over m'} \sum_{\vec{k}_1, \vec{k}_2, \vec{k}_3   \in \Psi} (\delta(\vec{k}_1)\delta(\vec{k}_2)\delta(\vec{k}_3))\delta_{Dirac}(\vec{k}_1+\vec{k}_2+\vec{k}_3),
\label{sumpowbisp}
\end{eqnarray}
where $\Phi$ is the set composed by all wavenumbers $\vec{k}_1$ (with
 $\vec{k}_2=-\vec{k}_1$) in the thin shell ($k_1 \pm \Delta k $,
 $\Delta k \ll k_1 $), and $\Psi$ is the set composed by all triangles
 with the same shapes formed by $\vec{k}_1$ and $\vec{k}_2$ (with
 $\vec{k}_3=-\vec{k}_2-\vec{k}_1$) in their thin shells respectively.
 $m$ and $m'$ are the numbers of the pairs and the triangles to be
 averaged.

Although it is very efficient to compute $\deltk$ with FFT, it is
 important to remember the numerical limitations caused by FFT. As
 shown in Jing (1992; 2004), the mass assignment onto a grid for FFT
 has two effects: the smoothing effect and the sampling effect. The
 smoothing effect has been considered by many authors in previous
 bistpectrum measurements (e.g. Scoccimarro et al. 1998, their
 Appendix), but the sampling effect has not. Because both effects are
 coupled, the smoothing effect cannot be fully corrected if the
 sampling effect is not considered. These effects must be taken
 into account for a precision analysis such as the current work.
 First, the particle distribution must be sampled at the FFT grid
 points. Here we adopt the Nearest-Grid-Point (NGP) (Efstathiou et
 al. 1985) approximation to assign the particle mass to the grid. This
 generally leads to a smoothing of the density field on the scale of
 $\sim \Delta x$ in the coordinate space. Higher order mass assignment
 schemes, e.g. cloud-in-cell (CIC) or triangular shaped cloud (TSC),
 cannot avoid the smoothing issue either; in fact they increase the
 smoothing even to a larger scale. In the next subsection, we will
 show that the mass assignment with NGP has the smoothing effect for
 $k>k_{Ny}/3$ which essentially limits the usable Fourier modes to
 $k<k_{Ny}/3$ .  For a 3D FFT with $N_m=1024$, $k$ is thus limited to
 $<160k_b$ that is not enough for fully exploring the non-linear
 properties. A much bigger $N_m$ would require a huge computer
 resource for the FFT computation. In the next section, we will show
 that we can overcome this problem effectively with a 2D FFT.

\subsection{Numerical effects on the power spectrum and bispectrum}
When measuring the power spectrum and bispectrum from a simulation we
must take account for the numerical artifacts, such as the
discreteness effect, the finite box size, and the force
softening. These limit the dynamical range of the simulation, and thus
affect the measured power spectrum and bispectrum. Other important
effect is that introduced by the FFT. Here we will address how to
correct and/or account for these numerical artifacts.
\subsubsection{Discreteness effects}
Since the power spectrum and bispectrum are measured from simulations with a
finite number of particles, we need to correct for the discreteness
effect arising from the Poisson shot noise. According to
Peebles (1980), we divide the volume $V_{\mu}$ into infinitesimal
elements $\{dV_i\}$ with $n_i$ objects inside $dV_i$.  The over-density
can be written as: $\delta_i=(n_i-\bar{n})/\bar{n}$ ($\bar{n}$ is the
mean number density), and its Fourier transformation can be expressed
as:
\begin{eqnarray}
\delta^d(\vec{k})={1 \over N}\sum_i n_i e^{i\vec{r}_i \cdot \vec{k}} -\delta_{Dirac}({\vec{k},0)},
\end{eqnarray}
where $N=\bar{n}V_{\mu}$, is the number of particles in
$V_{\mu}$. Since $dV_i$ is taken so small that $n_i$ is either 0 or 1,
we have $n_i=n_i^2=n_i^3=\cdots$.

For the power spectrum, we can find the ensemble average of
$\delta^d(\vec{k}_1) \delta^{d*}(\vec{k}_2)$ which reads:
\begin{eqnarray}
\langle \delta^d(\vec{k}_1) \delta^{d*}(\vec{k}_2)\rangle &=& {1 \over N^2}\sum_{i,j}\langle n_i n_j \rangle e^{i\vec{r}_i \cdot \vec{k}_1 -i\vec{r}_j \cdot \vec{k}_2}-\delta_{Dirac}(\vec{k}_1,0)\delta_{Dirac}(\vec{k}_2,0) \nonumber\\
 &=& \langle \delta(\vec{k}_1)\delta^*(\vec{k}_2) \rangle +{1 \over N} \delta_{Dirac}(\vec{k}_1,\vec{k}_2)\,.
\end{eqnarray}
Finally, we get the desired result:
\begin{equation}
P(k)=\langle|\delta^d(\vec{k})|^2\rangle -{1 \over N}.
\label{powedesctete}
\end{equation}

>From Eq.~(\ref{powedesctete}) we know that the discreteness (or shot noise) effect gives an additional term $1/N$ to the power spectrum. We can correct for the shot noise in the power spectrum easily.
In analogy with the power spectrum, we write the bispectrum :
\begin{eqnarray}
\langle \delta^d(\vec{k}_1)\delta^d(\vec{k}_2)\delta^d(\vec{k}_3) \rangle &=& 
\langle \delta(\vec{k}_1)\delta(\vec{k}_2)\delta(\vec{k}_3) \rangle\nonumber\\& & +{1 \over N}[P(\vec{k}_1)+P(\vec{k}_2)+ P(\vec{k}_3)]+{1 \over N^2},
\end{eqnarray}
The bispectrum with the shot noise removed can be expressed as:
\begin{eqnarray}
B(\vec{k}_1,\vec{k}_2,\vec{k}_3) &=&\langle \delta^d(\vec{k}_1)\delta^d(\vec{k}_2)\delta^d(\vec{k}_3) \rangle \nonumber\\ & & -{1 \over N}[P(\vec{k}_1)+P(\vec{k}_2)+ P(\vec{k}_3)]-{1 \over N^2}.
\end{eqnarray}

\subsubsection{Force softening and box size}
 In the simulation, in order to suppress two-body encounters, a
 softening must be applied when calculating the gravitational
 interaction.  This induces an error in the integration of particle
 trajectories at small scale. We must impose some constraint on the
 scale below which the numerical effect dominates the clustering in
 the simulation. The cutoff can be a few times of the softening
 length.

The box size of the simulation is finite, so there is no clustering
power beyond the simulation box. On one hand, this results in a
limited number of Fourier modes at $k\ga k_b$ which can influence the
accuracy of measuring the clustering spectra. On the other hand, this
large-scale cutoff may also affect the clustering on scale much smaller
than the box size, because the coupling between different scales can
be important on non-linear and quasilinear scales.

Both the force softening and the box-size cutoff are expected to break
down the scaling property of the self-similar evolution. We will use
the expected scaling to quantify these numerical artifacts.

\subsubsection{Mass assignment}
When doing the FFT, we first need to collect density values on grids
(usually called mass assignment). The mass assignment in fact is
equivalent to convolving the density field by one chosen function
$W(r)$ and sampling the convolved density on a finite number of grid
points, therefore the FFT of $\rho(\vec{r}_g)$ generally is not equal
to the FT of $\rho(\vec{r})$. In this work, we have adopted the NGP
(the nearest grid point) scheme to assign particles to the mesh. The
finite sampling of the convolved density results in the summation of
the aliased power spectrum or bispectrum.

We can correct the alias effect on the power spectrum through a
theoretical calculation (Jing 1992, 2004; Baugh \& Efstathiou 1994;
Smith et al. 2003).  But it is complicated to correct for the alias
effect on the bispectrum through a theoretical
calculation. Fortunately, we find that the 2D statistical properties
are identical to the 3D statistical properties in Fourier space (\S
5). Thus we can make use of the 2D density field instead of the 3D
density field when calculating the bispectrum at small scales, because
the 2D FFT needs less computer memory than the 3D case. In fact, the
larger number of grid points $N_m$ for the FFT, the smaller scale
where the alias effect takes place. The 2D FFT can overcome the
limitation of the computation and involve little alias effect. Thus we
can extend the measurement of the bispectrum to very small scales with
little numerical artifact.
 
\section{The method}

Let $\delta_{3D}(\vec{r})$ be the over-density in 3D real space, and
$\delta_{2D}(\vec{x})$ the over-density in 2D real space, 
\begin{eqnarray}
\delta_{3D}(\vec{r}) &\equiv& \frac{\rhor - \bar\rho_0}{\bar\rho_0}, \nonumber \\
\delta_{2D}(\vec{x}) &\equiv& \frac{\omega(\vec{x}) - \bar\omega_0}{\bar\omega_0}.
\label{delta23}
\end{eqnarray}
where $\rho(\vec{r})$ and $\omega(\vec{x})$ are the density field of
the 3D real space and the 2D real space respectively, $\bar\rho_0$ and
$\bar\omega_0$ are the mean density correspondingly. The 2D density
field is defined by integrating the 3D density field
along one direction, say $z$-axis, 
\begin{eqnarray}
\omega(\vec{x}) \equiv \int_L \rho(\vec{r}) dz  \; , \; 
\bar\omega_0 \equiv \bar\rho_0 L\,.
\label{rhointx}
\end{eqnarray}
The over-density in 2D space is:
\begin{eqnarray}
\delta_{2D}(\vec{x}) &=& \frac{\omega(\vec{x})-\bar\omega_0}{\bar\omega_0}={1\over L}\int_L \delta_{3D}(\vec{r})\, dz .
\label{delta2Dx}
\end{eqnarray}
Eq.~(\ref{deltak1}) gives the Fourier transformation of the
over-density in 3D space. The Fourier transformation of the
over-density in 2D space can be expressed as
\begin{equation}
\delta_{2D}(\vec{k}_{2D}) = {1\over A} \int_A  \delta_{2D}(\vec{x})e^{i \vec{x} \cdot \vec{k}_{2D}}\, d\vec{x},
\label{delta2Dk} 
\end{equation}
where A is the normalization surface area. Inserting Eq.~(\ref{delta2Dx}) into Eq.~(\ref{delta2Dk}), we get
\begin{eqnarray}
\delta_{2D}(\vec{k}_{2D}) &=& {1\over A}\int_A{1\over L}\int_L\delta_{3D}(\vec{r})e^{i\vec{x}\cdot \vec{k}_{2D}} \,dzd\vec{x}\nonumber\\
 &=& {1\over V_\mu}\int_A \int_L \sum_{\vec{k^\prime}} \delta_{3D}(\vec{k^\prime})e^{-i\vec{r} \cdot \vec{k^\prime}} e^{i\vec{x}\cdot \vec{k}_{2D}}\,dzd\vec{x}\nonumber\\
 &=& \sum_{\vec{k^\prime}_{2D},k^\prime_z} \delta_{3D}(\vec{k^\prime}_{2D},k^\prime_z)\delta_{Dirac}(\vec{k^\prime}_{2D},\vec{k}_{2D}) \delta_{Dirac}(k_z^\prime,0) \nonumber\\
 &=& \delta_{3D}(\vec{k}_{2D},0).
\label{delta2D3Dk}
\end{eqnarray}
Eq.~(\ref{delta2D3Dk}) gives an important hint that we can connect the statistical properties of 3D Fourier space with that of 2D Fourier space. Following Eq.~(\ref{powbisp}), we get
\begin{eqnarray}
\langle \delta_{2D}(\vec{k}_1)\delta_{2D}(\vec{k}_2) \rangle&=&\langle \delta_{3D}(\vec{k}_1,0)\delta_{3D}(\vec{k}_2,0) \rangle, \nonumber\\ 
\langle \delta_{2D}(\vec{k}_1)\delta_{2D}(\vec{k}_2)\delta_{2D}(\vec{k}_3) \rangle &=& \langle \delta_{3D}(\vec{k}_1,0)\delta_{3D}(\vec{k}_2,0)\delta_{3D}(\vec{k}_3,0) \rangle .
\label{delta2d3d0} 
\end{eqnarray}
Consistent with the Cosmological Principle (the isotropic property of
the Universe), Eq.~(\ref{delta2d3d0}) means that the 2D density field
in Fourier space has the same statistical properties as the 3D field.

\begin{figure}
\hbox{\hspace{0.5cm}\psfig{figure=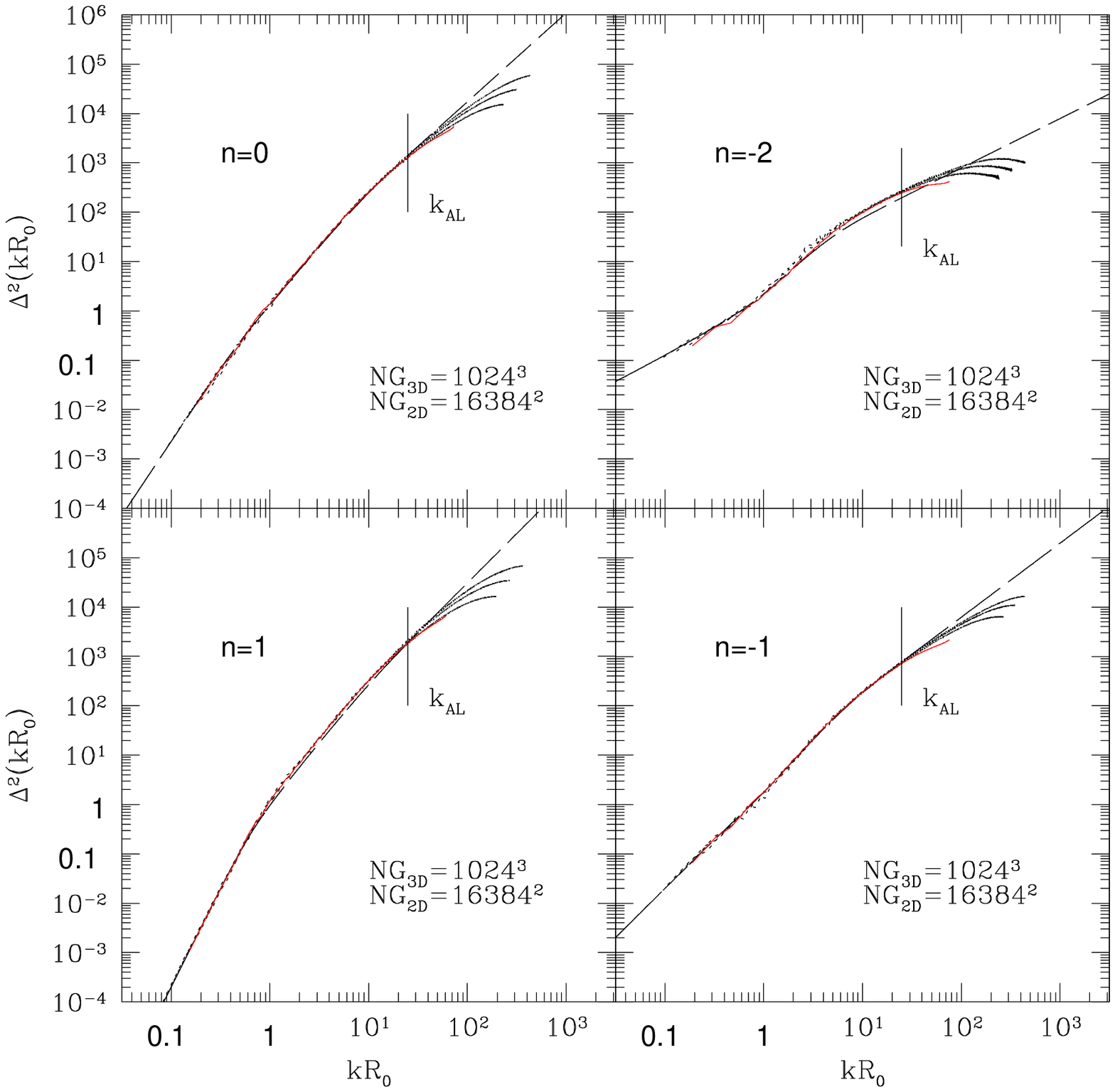,width=14.cm}}
\caption{Comparison of the power spectrum measured from the 2D and 3D
density fields. The data are the last three outputs of the simulations
with the spectral index $n=1, 0, -1, -2$. In each panel, the three
short-dashed lines (middle) represent the 2D power spectrum measured
from the last three outputs of the simulation, the solid line (bottom)
is the 3D power spectrum measured from the last output of the
simulation, and the long-dashed line (up) is the prediction by the
fitting formula of Peacock \& Dodds (1996). The value of $\rm NG $
represents the number of grid points adopted in their FFT
respectively. We plot $\Delta^2(kR_0)$ ($ k=2\pi n /L, n=1,2 \cdots$.)
until $n=512$ for the 3D power spectrum, and until $n=3000$ for the 2D
power spectrum. The wavenumber $k_{AL}$ at the vertical line is
$k_{Ny}/3$ for 3D power spectrum. $R_0$ is the characteristic scale
defined in Eq.~(\ref{characterR0}).}
\label{figpow2D3D}
\end{figure}

In Figure \ref{figpow2D3D} we compare the 2D power spectrum with the
 3D power spectrum measured from the simulations with initial spectral
 index $1$, $0$, $-1$, and $-2$. The number of grid points of the 2D
 FFT is $16384^2$, and the number of the 3D FFT is $1024^3$. For the
 3D power spectrum we plot $\Delta^2(kR_0)$ ($k=n k_b$, $n=1,2
 \cdots$) for the last output of the simulations until $n=512$.  As to
 the 2D case, we measured the power spectrum with the last three
 outputs of the simulations under the scaling transformation, and plot
 $\Delta^2(kR_0)$ ( $k=2 \pi n/L, n=1,2$ $\cdots$) until $n=3000$.  Here we
 do not correct for the alias effect, and find that the alias begins
 to affect the power spectrum when the wavenumber is larger than
 $k_{AL}=k_{Ny}/3$ as indicated by the vertical lines in the
 picture. This picture also shows that the 2D power spectrum is
 identical to the 3D power spectrum when the alias effect is
 negligible.

\begin{figure}
\hbox{\hspace{0.5cm}\psfig{figure=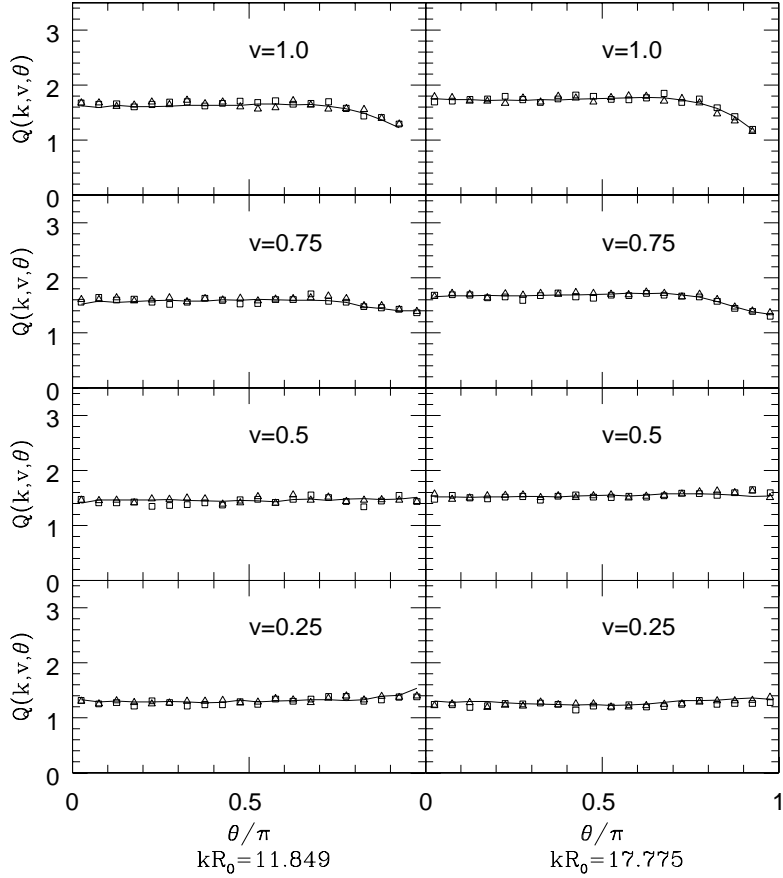,width=14.cm}}
\caption{Comparison of the reduced bispectrum measured from the 2D and
the 3D density fields. The data are the last two outputs ($a=0.792$,
$a=1.0$) of the simulation with spectral index $n=-1$. The solid line
corresponds to the 3D reduced bispectrum at the last output of the
simulation, and the open symbols are for the 2D reduced bispectrum at
the last two outputs of the simulation under the scaling
transformation. The left panels give $Q(k,v,\theta)$ for $kR_0=11.8$,
and the right ones for $kR_0=17.8$.}
\label{figQ_test23D}
\end{figure}

In Figure~\ref{figQ_test23D} we compare the reduced bispectrum
measured in the 2D and 3D density fields with the simulation of
$n=-1$. We express the reduced bispectrum as a function of $k$, $v$
and $\theta$ as defined in Eq.~(\ref{shapetriangle}). Each panel shows
$Q(k,v,\theta)$ as a function of the angle $\theta$ between
$\vec{k}_1$ and $\vec{k}_2$ with different $k$ and $v$.  For the 3D
bispectrum we only show the results of the last output for this
simulation ($a=1$). For the 2D case we get the reduced bispectrum with
the last two outputs of the simulation ($a=0.792,1$) under the scaling
transformation. The corresponding $k$ values of the triangles are less
than $k_{AL}$, so the alias does not affect the measured results. The
figure shows that the bispectrum measured in 2D agrees very well with
that in 3D, as expected.

From Figures~\ref{figpow2D3D} and \ref{figQ_test23D}, we conclude that
it is reliable to use the 2D FFT instead of the 3D FFT to calculate
the power spectrum and bispectrum on small scales. For the 2D particle
distribution, an FFT with $N_m=8192$ requires computer memory about
0.3 Gbyte, even that with $N_m=16384$ requires memory about 1 Gbyte
only. Therefore we can extend the bispectrum measurement into the
strong nonlinear regime with little computer limitation.

\begin{figure}
\hbox{\hspace{0.5cm}\psfig{figure=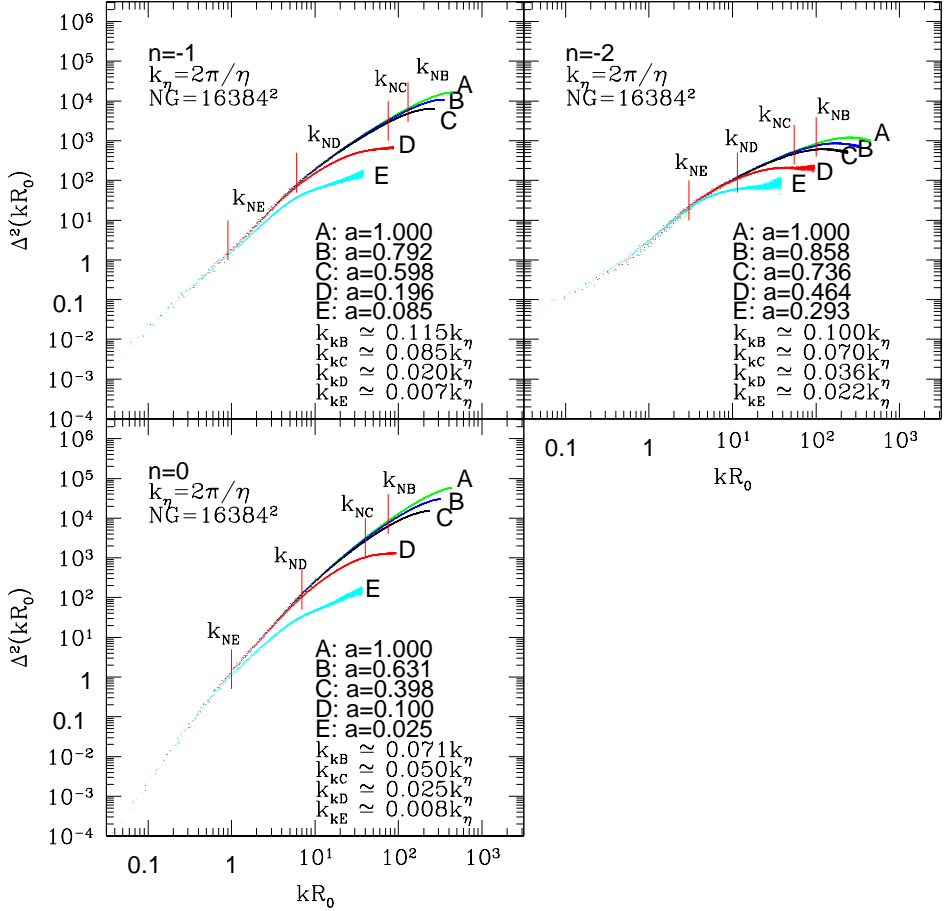,width=14.cm}}
\caption{ The power spectrum measured by FFT with the same number of
grid points ($NG=16384^2$) for five epochs of the simulations, and
scaled by the characteristic scale. These results are plotted by the
lines A, B, C, D, E in each panel until $k_{AL}$, and their epochs are
shown in the picture. The numerical artifacts begin to affect the
power spectrum at the vertical line indicated by $k_{NB}$, $k_{NC}$,
$k_{ND}$, $k_{NE}$ (in units of $k_{\eta}$, $\eta$ is
the softening length), at which
the deviation of the power spectrum from the true power spectrum 
is about 10 percent.
}
\label{softening}
\end{figure}

\begin{figure}
\hbox{\hspace{0.5cm}\psfig{figure=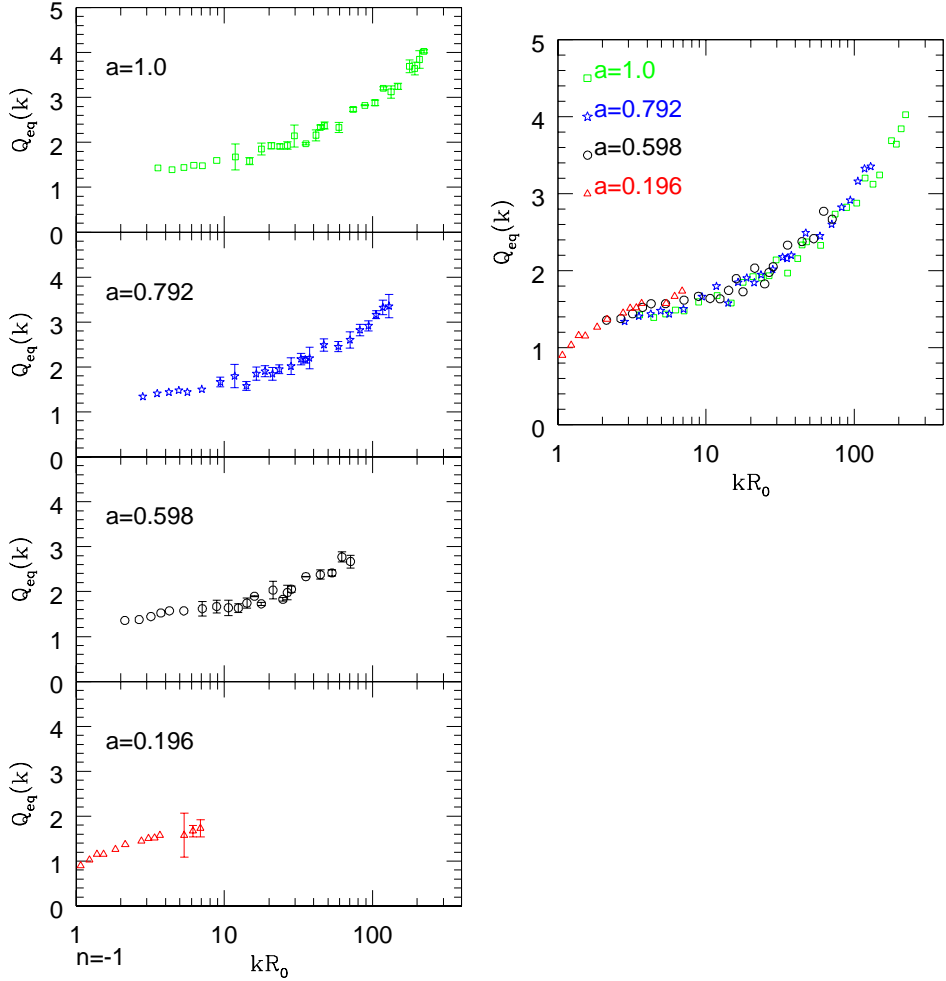,width=14.cm}}
\caption{The reduced bispectrum for equilateral triangles measured by
 FFT with $NG_{2D}=8192^2$ or $NG_{3D}=1024^3$ for four epochs of the
 $n=-1$ simulation. The results are plotted up to the wavenumber at
 which the softening begins to influence the power spectrum (the
 vertical lines in Figure \ref{softening}).  Each panel on the left
 shows the results for one epoch. The points with and without error
 bars are measured from the 2D and 3D density fields respectively. We
 compare these results on the right panel where for clarity we do not
 show the errors.}
\label{Qsoft2920}
\end{figure}

 In fact, $N_m=8192$ is sufficient to avoid the alias effect. In
 Figure~\ref{softening}, we show the power spectra at different epochs
 measured with the same number of FFT grid points. The deviation of
 the power spectrum from the scaling expectation at $k<k_{AL}$ should
 be attributed to the numerical artifacts in the simulations. From the
 figure, we find that these artifacts begin to influence the power
 spectrum at $k\approx 0.1 k_{\eta}$ for late outputs where
 $k_{\eta}=2\pi /\eta$. We believe this is mainly caused by the force
 softening. For early outputs, the deviation from the scaling happens
 at a larger scale, which we attribute to the cutoff of the initial
 fluctuation at $k>k_{Ny}$ as well as the initial distribution of the
 particles on grid. All this shows that $N_m=8192$ is sufficient to
 explore the non-linear features in our simulations.  In Figure
 ~\ref{Qsoft2920} we check the softening effect on the reduced
 bispectrum.  For simplicity we give the results only for spectral
 index $-1$.  The left panels show the reduced bispectrum for
 equilateral triangles at four epochs up to the wavenumber where the
 softening begins to affect the power spectrum (as defined in Figure
 3). The points with error bars are measured from the 2D density
 field, and those without error bars are obtained from the 3D density
 field. The error bars of the 2D bispectra are estimated from three
 projections. We compare these results on the right panel using the
 similarity scaling, which shows that the reduced bispectrum is not
 affected by the softening for the wavenumbers less than the softening
 limit set by the power spectrum analysis (Figure 3). We will use this
 criteria to minimize the softening effect in our $Q$ measurement.

We should point out that for a fixed range of $k$, the number of the
Fourier modes is much smaller in 2D than in 3D. This limitation
implies that the 2D FFT is not appropriate for the measurement at
small wavenumbers. In our current analysis, we do both the 3D FFT with
$N_m=1024$ and the 2D FFT with $N_m=8192$ for each simulation. For the
wavenumber $k<2\pi/L\times 512/3$, we use $\deltk$ based on the 3D FFT;
otherwise we use $\deltk$ based on the 2D FFT.  For the 2D case, we
have projected the density field along three axes. The 2D bispectrum is
measured by averaging over the results of the three projections.


\section{Results}
\subsection{Numerical results}
\begin{figure}
\hbox{\hspace{0.5cm}\psfig{figure=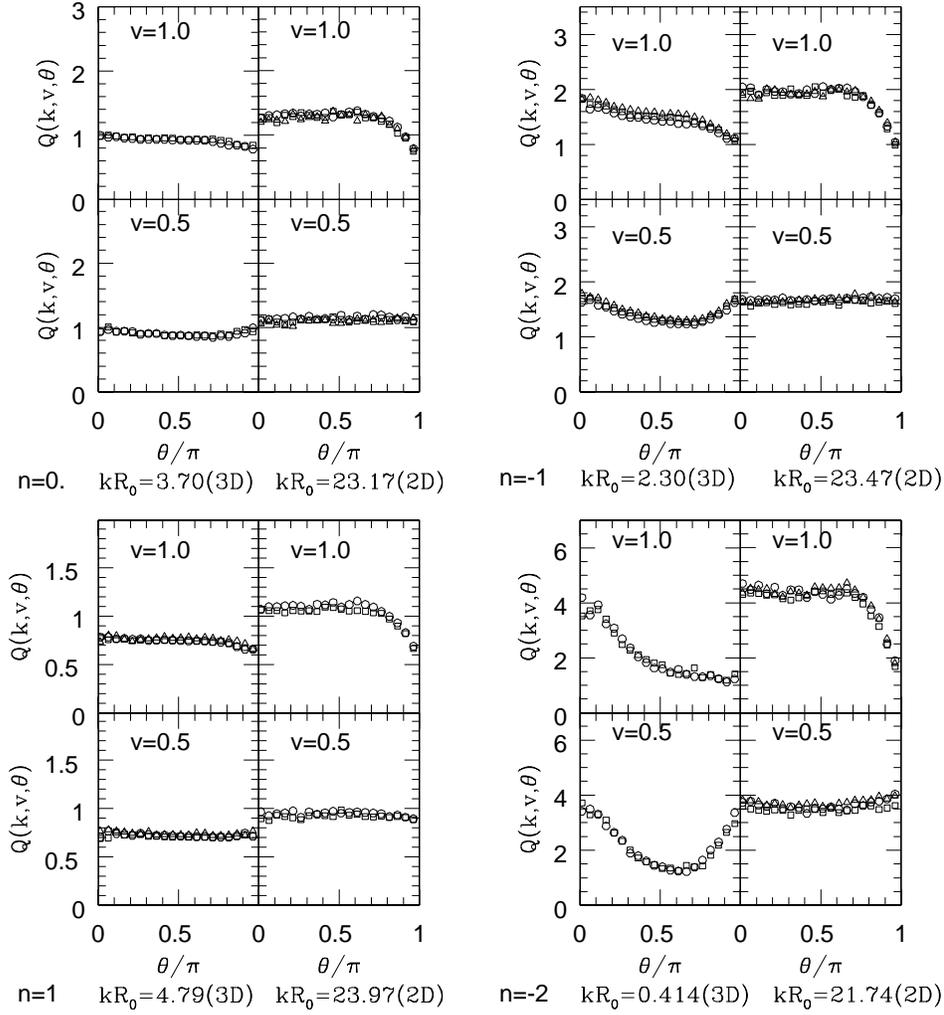,width=14.cm}}
\caption{The reduced bispectrum $Q(k,v,\theta)$ as a function of the
angle $\theta$ between $\vec{k}_1$ and $\vec{k}_2$ for two or three
outputs scaled by the characteristic scale. These results are measured
by 3D FFT or 2D FFT as indicated at the bottom. The scales are also
indicated. Different symbols (open circles, open triangles, open
squares) stand for the results at different outputs.}
\label{figQ_scaled}
\end{figure}

In this section, we show the reduced bispectrum from the quasilinear
regime to the strong non-linear regime measured with our new method.
We have measured the bispectrum for many triangle shapes at different
scales. In each case, we scale the results of two or three outputs
with the characteristic scale defined by Eq. (\ref{characterR0}) in
order to make sure that these results are not affected by numerical
effects. For simplicity, in Figure~\ref{figQ_scaled} we show only a
few measured bispectra for spectral index $1$, $0$, $-1$, $-2$, and
these results scale very well.  From the measured results (Figures
~\ref{fitting2910}, \ref{fitting2920} and \ref{fitting2930}; left
columns), we find that the results are in good agreement with the
one-loop PT prediction in the quasilinear regime (Scoccimarro et
al. 1998): the reduced bispectrum $Q$ is higher than the second-order
PT for $n=-2$ and is smaller for $n\ge -1$.  We also find that $Q$
depends on the triangle shape and size even in the strong nonlinear
regime, which is in contrast with the hypothesis adopted by SF99 that
$Q$ is a constant in this regime. It decreases with the initial
spectral index and increases with the wavenumber, similar to what
found for cold dark matter models (Jing \& B\"orner 1998, Ma \& Fry
2000, Scoccimarro, et al. 2001b, Takada \& Jain 2003a).

\subsection{A fitting formula for the bispectrum}
SF99 presented a fitting formula for the bispectrum in scale-free
clustering models. From Figures~\ref{fitting2910}, \ref{fitting2920}
and \ref{fitting2930} we find that the reduced bispectrum
predicted by this formula agrees with the measured bispectrum at
linear and quasilinear scales, and works well for some triangles with
special shapes (such as $k_1=2k_2$ in the $n=-1$ model) in the strong
nonlinear regime. But this formula cannot generally follow the
bispectrum accurately at strongly nonlinear scales, because they assumed
that the normalized bispectrum Q is a constant for a given initial
spectral index, independent of the triangle's wavenumber and shape. We
provide a new fitting formula in this section to describe the
nonlinear evolution of the bispectrum.

\begin{figure}
\hbox{\hspace{0.5cm}\psfig{figure=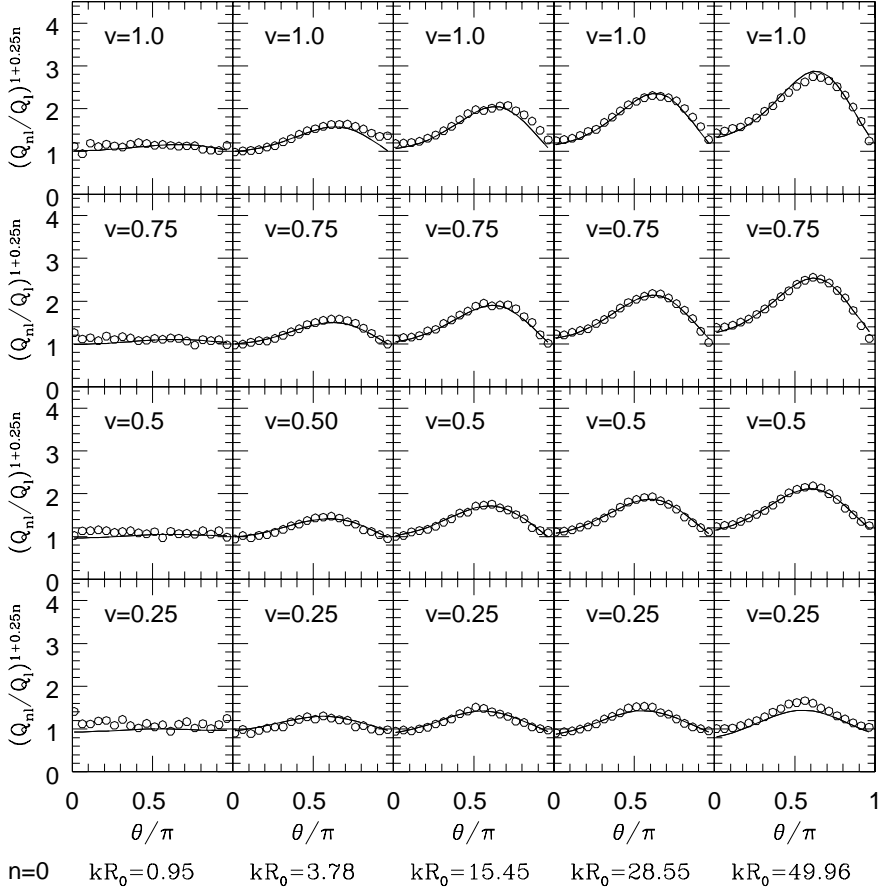,width=14.cm}}
\caption{The value of $(Q_{nl}/Q_l)^{1/(1+0.25n)}$ as a function of
$\theta$ for the spectral index $n=0$. $Q_{nl}$ is the reduced
bispectrum measured from the simulation, and $Q_l$ is the weakly
non-linear reduced bispectrum predicted by PT2. The solid line shows
our fitting formula for $(Q_{nl}/Q_l)^{1/(1+0.25n)}$, and the circle
symbols are the simulation results. Vertical panel columns have the
same $kR_0$, as indicated at the bottom.}
\label{ratio2910}
\end{figure}

\begin{figure}
\hbox{\hspace{0.5cm}\psfig{figure=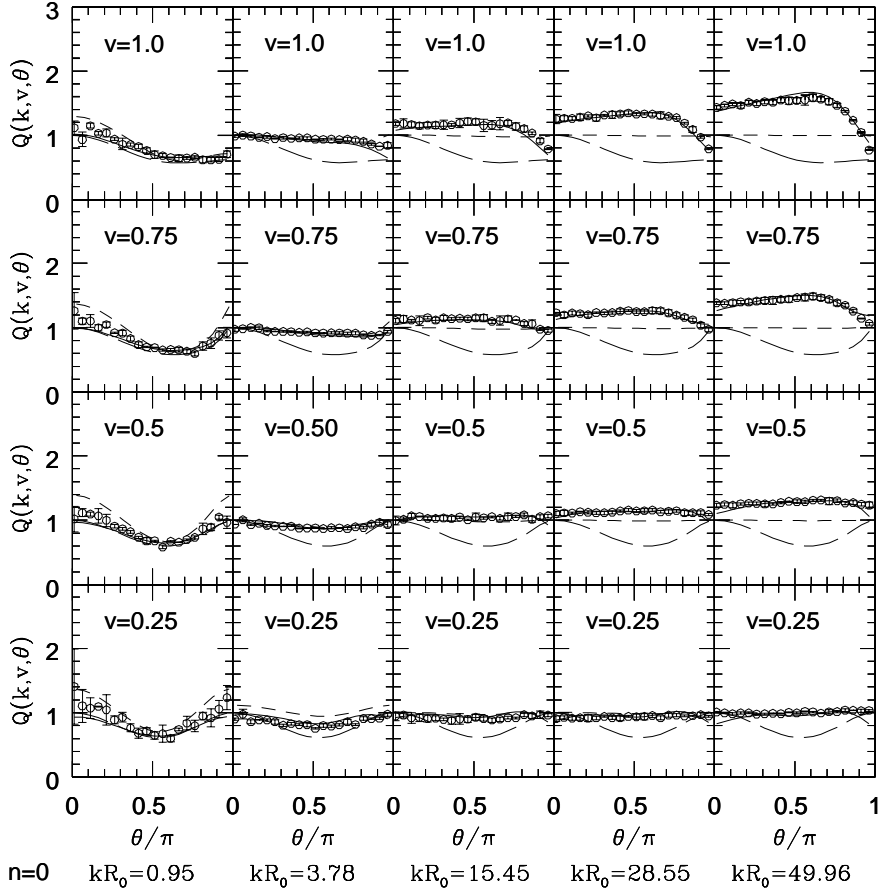,width=14.cm}}
\caption{The reduced bispectrum $Q(k,v,\theta)$ measured from the
simulation with the spectral index $n=0$ (open circles), compared with
the predictions by PT2 (long-dashed lines), by the fitting formula of
SF99 (short-dashed lines), and by the fitting formula in this paper
(solid lines). Vertical panel columns have the same $kR_0$, as
indicated at the bottom. }
\label{fitting2910}
\end{figure}

\begin{figure}
\hbox{\hspace{0.5cm}\psfig{figure=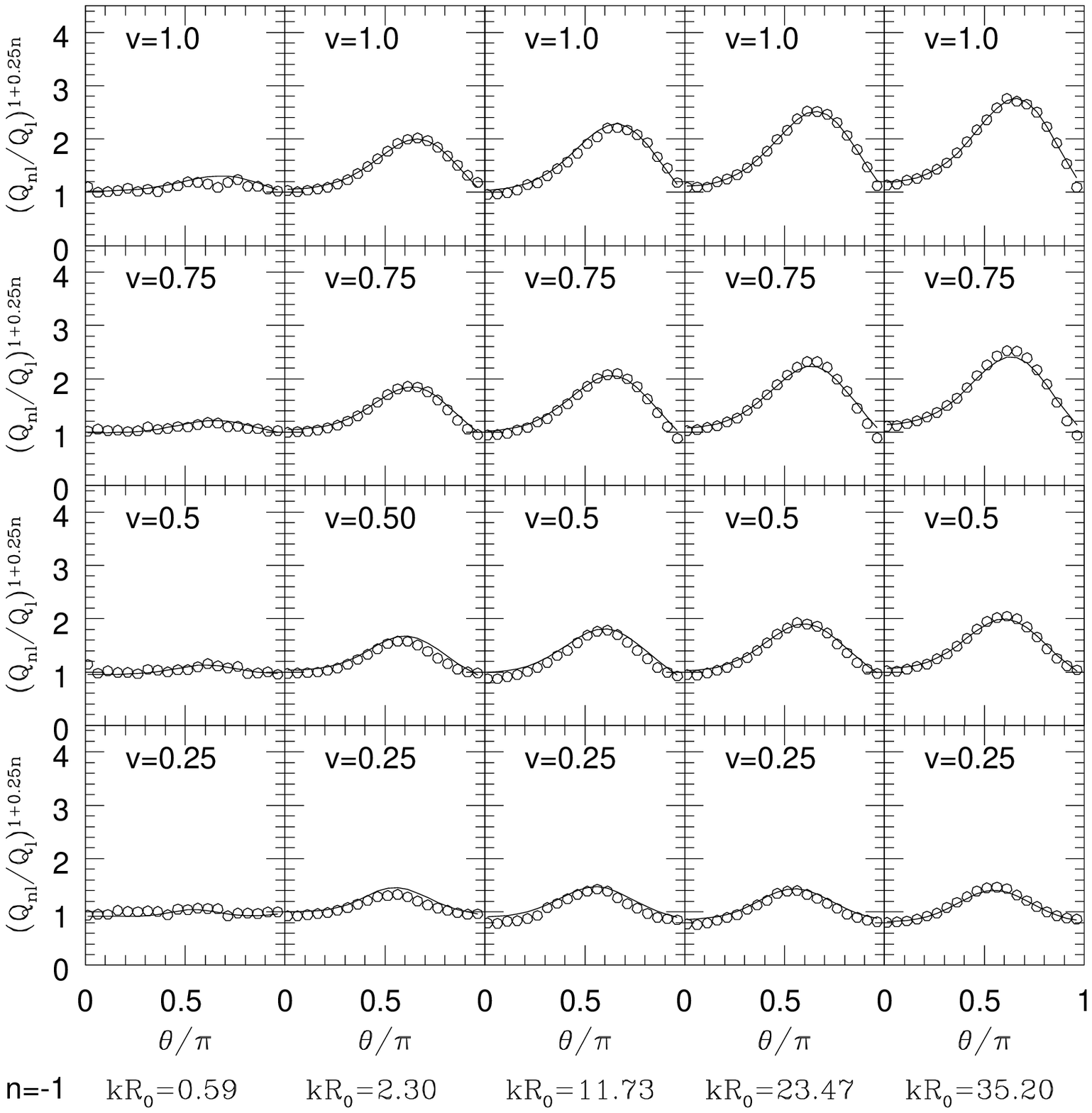,width=14.cm}}
\caption{Same as Fig.\ref{ratio2910}, but for the spectral index $n=-1$.}
\label{ratio2920}
\end{figure}

\begin{figure}
\hbox{\hspace{0.5cm}\psfig{figure=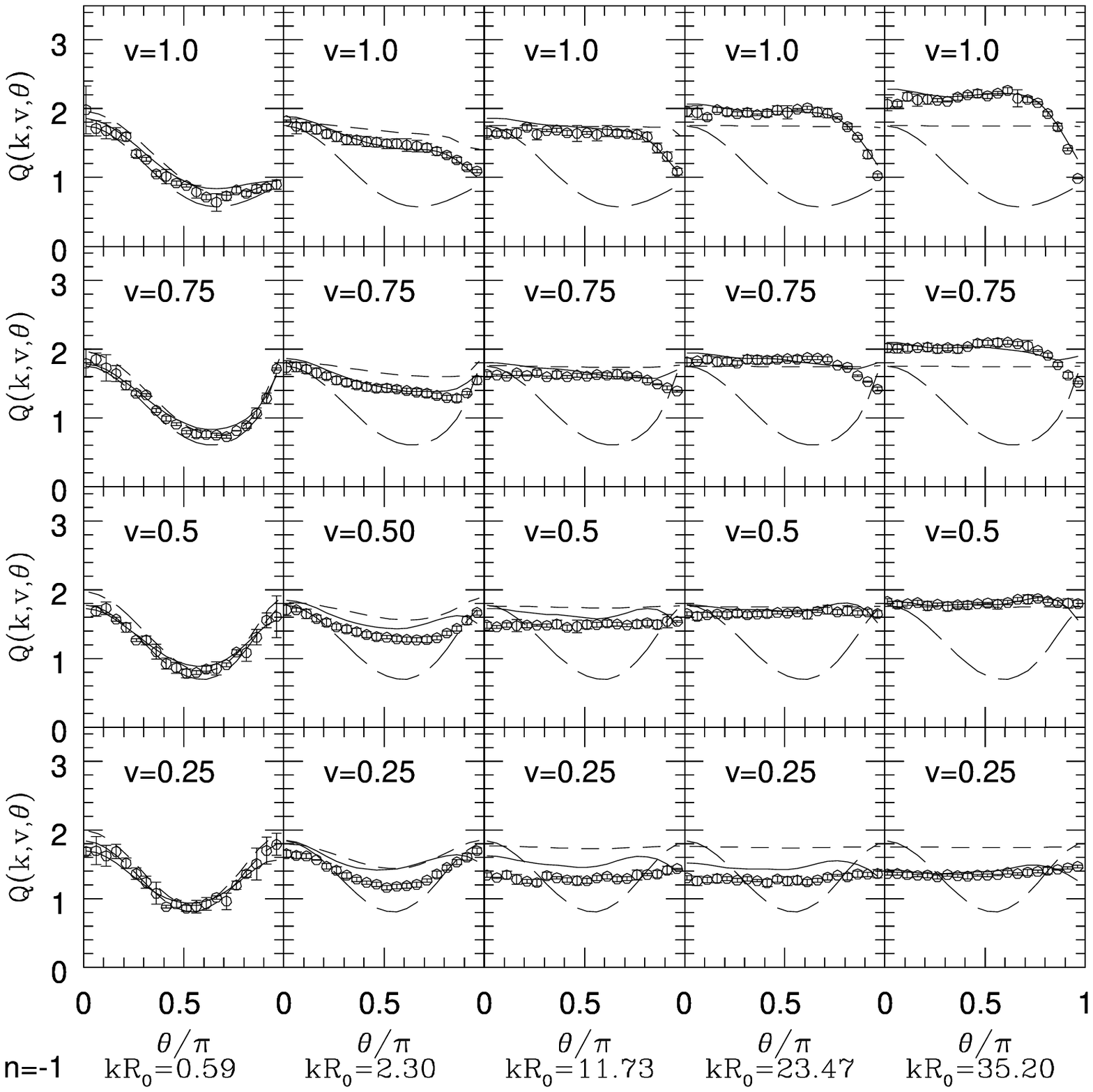,width=14.cm}}
\caption{Same as Fig.\ref{fitting2910}, but for the spectral index $n=-1$.}
\label{fitting2920}
\end{figure}

\begin{figure}
\hbox{\hspace{0.5cm}\psfig{figure=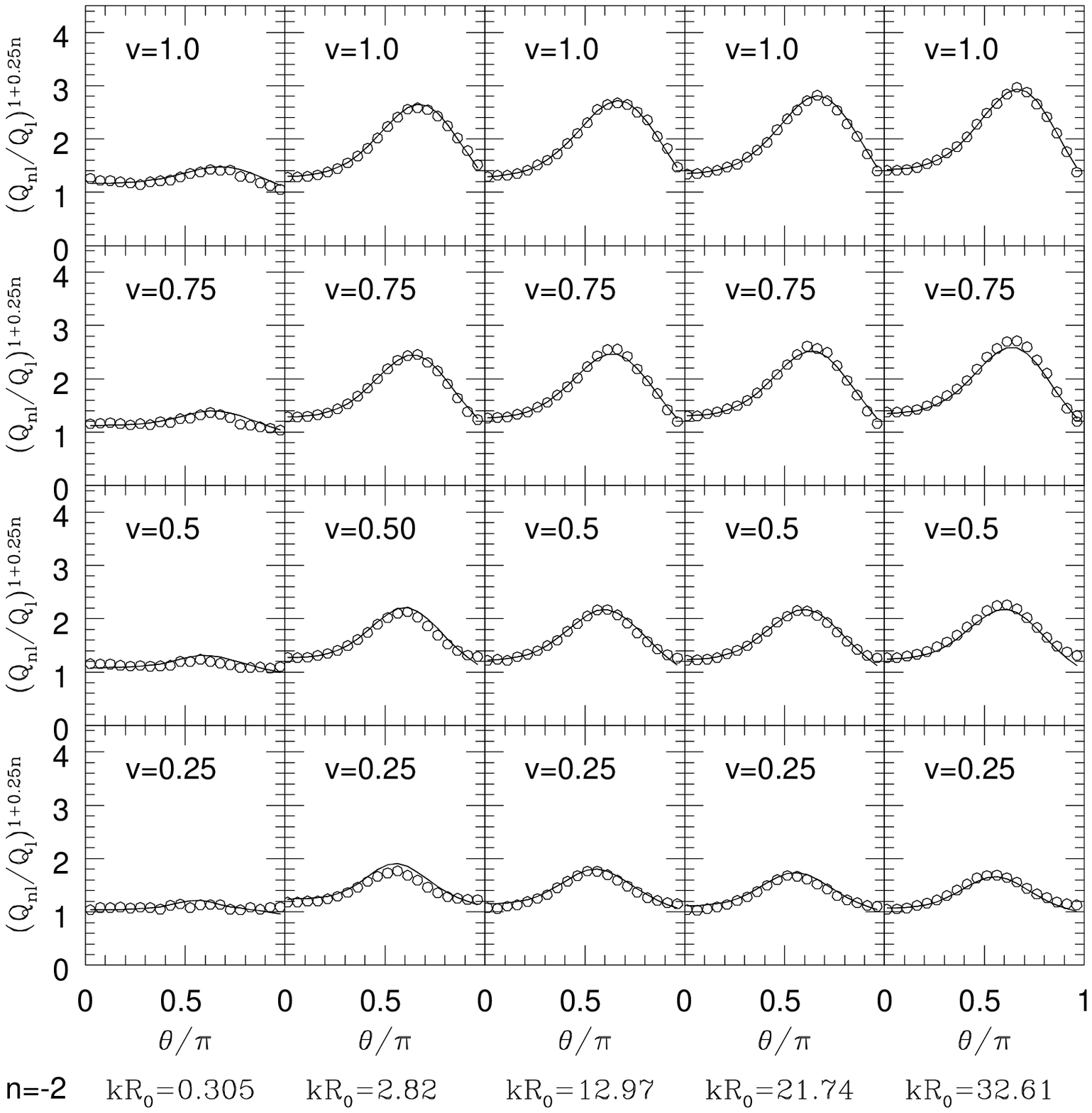,width=14.cm}}
\caption{Same as Fig.\ref{ratio2910}, but for the spectral index $n=-2$.}
\label{ratio2930}
\end{figure}

\begin{figure}
\hbox{\hspace{0.5cm}\psfig{figure=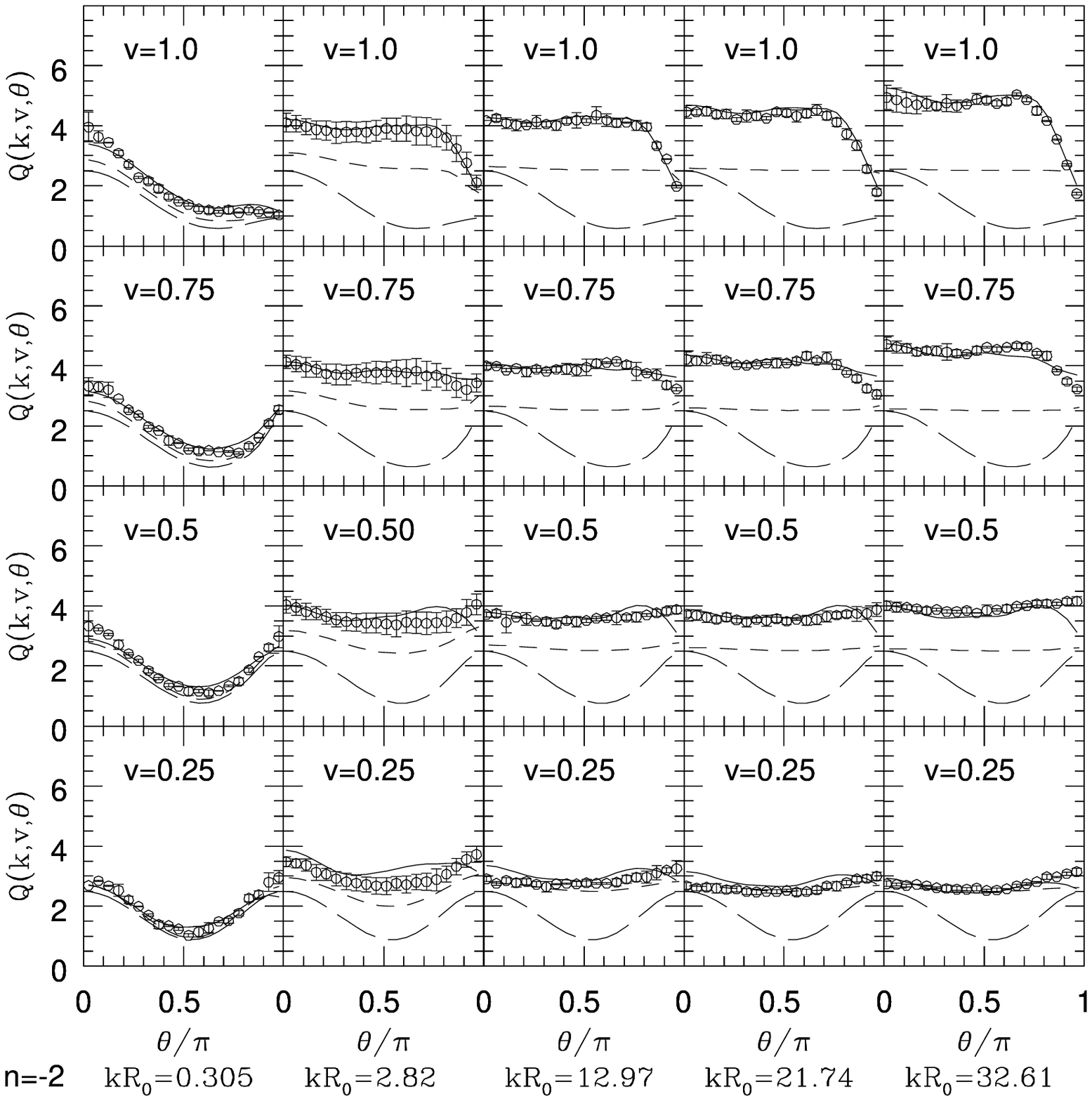,width=14.cm}}
\caption{Same as Fig.\ref{fitting2910}, but for the spectral index $n=-2$.}
\label{fitting2930}
\end{figure}

Hamilton et al. (1991) proposed a universal empirical relation between
the linear $\xi_L$ and non-linear $\xi_{NL}$ two-point correlation
functions. The relation is a powerful tool for predicting $\xi_{NL}$
in cosmological models.  Later, Jain, Mo, \& White (1995) demonstrated
that the relation of Hamilton et al. fails for the scale-free model of
$n=-2$. Peacock \& Dodds (1996) examined a large set of scale-free
models and CDM models, and obtained a accurate fitting formula for the
power spectrum which agrees with the results of N-body
simulations. Motivated by their work, we find that the ratio of the
measured bispectrum to the weakly non-linear bispectrum (the
second-order PT) has interesting properties, especially the behavior
of $(Q_{nl}/Q_l)^{1+0.25n}$ ($-2 \leq n \leq 0 $). A few typical
examples of $(Q_{nl}/Q_l)^{1+0.25n}$ are shown in
Figures~\ref{ratio2910}, \ref{ratio2920} and \ref{ratio2930} (the open
symbols) for the three models with $n=0$, $-1$, and $-2$. From these
results we expect that the relation between the weakly non-linear
bispectrum and the nonlinear bispectrum can be expressed as:
\begin{equation}
Q_{nl}(kR_0,v,\theta)=f_{nl}^{1/(1+0.25n)} \left(kR_0,v,\theta,n)
Q_l(kR_0,v,\theta)\right ),
\label{funcQnlQl}
\end{equation}
where $n$ satisfies $-2 \leq n \leq 0$.  At linear scales $Q_{nl}
\sim Q_l$, and $f_{nl}$ takes an asymptotic form as
$f_{nl}(kR_0,v,\theta,n) \sim 1 $ in the linear regime. From the 
measured $(Q_{nl}/Q_l)^{1+0.25n}$, we can find that 
$(Q_{nl}/Q_l)^{1+0.25n}$ as
a function of $\theta$ is approximated by a Gaussian function. Its
difference from the Gaussian function is dependent on the triangle shape,
scale and spectral index $n$. Taking into account of these factors, we
propose the following form for $f_{nl}(kR_0,v,\theta,n)$:
\begin{equation}
f_{nl}(kR_0,v,\theta,n ) ={ a_1(kR_0,v,n) \exp \left (-\displaystyle{( \theta/\pi - a_2(kR_0,v,n))^2 \over a_3(kR_0,v,n)} \right ) + a_4(kR_0,v,n) \over 1+a_5(kR_0,v,n)(\theta/\pi)^2 +a_6(kR_0,v,n)(\theta/\pi)^4},
\label{fittingfnl}
\end{equation} 
where $a_i(kR_0,v,n)$ are:
\begin{eqnarray}
a_1(kR_0,v,n)&=& \{1-{2 \over \exp(0.5(1-n)kR_0)+1}+[(4+n)0.002v+\nonumber\\& &(0.012+0.008n)]kR_0 \} [(0.1-0.3n)v+0.4],\nonumber\\
a_2(kR_0,v,n)&=&0.5+0.2v, \nonumber\\
a_3(kR_0,v,n)&=& [1.-{2 \over \exp(0.5kR_0)+1}+(0.01nv+0.001)kR_0]0.04+\nonumber\\& &0.06v+0.1^{(1-n)}, \nonumber\\
a_4(kR_0,v,n) &=& \{1.1+[(n^2)^{1.3}0.05-0.2]\tanh(2kR_0)+0.15\exp(-(0.3kR_0)^2)+\nonumber\\
& &(0.01-{0.05-0.035v\over 4v+2})kR_0\}/[1+(0.4-0.2v)^{1+kR_0}], \nonumber\\
a_6(kR_0,v,n) &=& [0.8-{2 \over \exp((0.5-n)kR_0)+1}+(0.01-0.005/(v+0.1))kR_0]  \nonumber\\
& & [2-{0.3 \over v+0.1}]+0.2n\tanh(2kR_0-1), \nonumber\\
a_5(kR_0,v,n) &=& -0.7a_6(kR_0,v,n),
\end{eqnarray}
where $\tanh(x)=(\exp(x)-\exp(-x))/ (\exp(x)+\exp(-x))$ is the hyperbolic 
function.

We have obtained the best-fitting parameters in $a_i(kR_0,v,n)$ by doing
a $\chi^2$ minimization between the predicted and measured bispectrum
for all triangles.  In Figures~\ref{ratio2910}, \ref{ratio2920} and
\ref{ratio2930} we compare the best fitting function $f_{nl}$ with our
measured $(Q_{nl}/Q_l)^{1+0.25n}$. The figures show that the fitting
formula (\ref{fittingfnl}) works very well for $-2 \leq n \leq 0$, for all
triangle shapes, and for all scales (characterized by $kR_0$) studied.
With the function $f_{nl}$, we can convert the weakly non-linear bispectrum into
the non-linear bispectrum. In Figures~\ref{fitting2910},
\ref{fitting2920} and \ref{fitting2930} we compare our measured
reduced bispectra with the prediction of our fitting formula. There we
also plot the prediction of the fitting formula of SF99. From the
figures, one can see that the fitting formula obtained in this paper
can accurately match the simulation results, much better than the
formula of SF99.

\begin{figure}
\hbox{\hspace{0.5cm}\psfig{figure=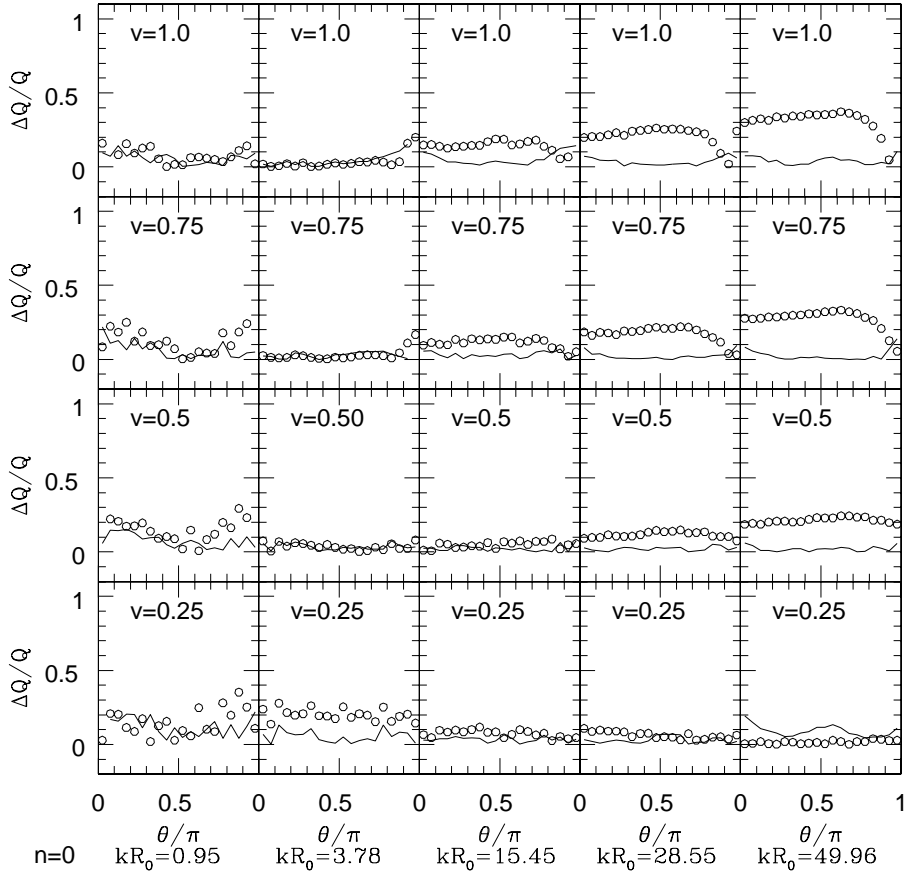,width=14.cm}}
\caption{The relative accuracy of the fitting formulae for the
bispectrum for the spectral index $n=0$. The open circles are for the
formula of SF99, and the solid lines are for the formula obtained in this
paper.}
\label{erro2910}
\end{figure}

\begin{figure}
\hbox{\hspace{0.5cm}\psfig{figure=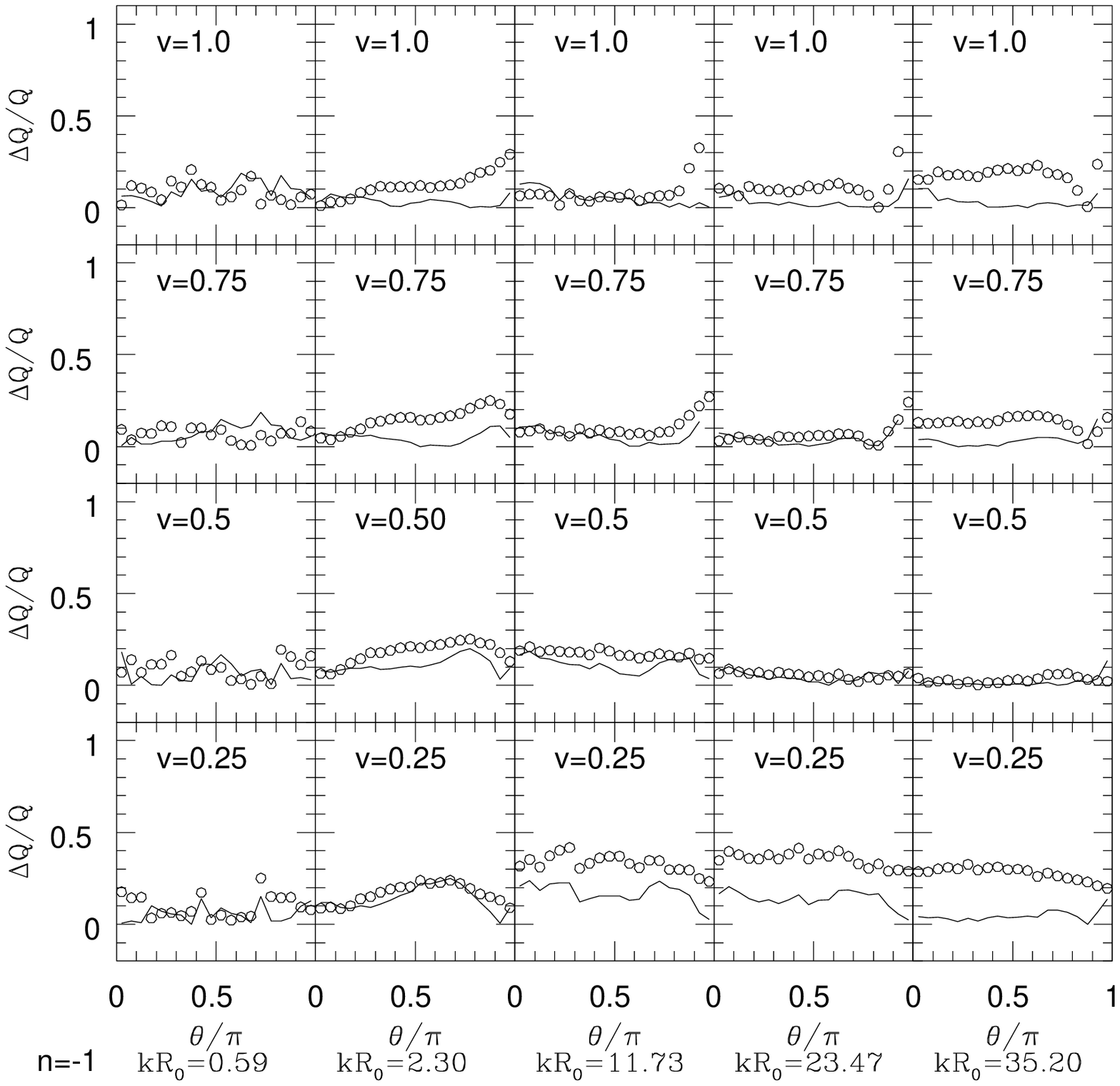,width=14.cm}}
\caption{Same as Fig.\ref{erro2910}, but for the spectral index $n=-1$.}
\label{erro2920}
\end{figure}

\begin{figure}
\hbox{\hspace{0.5cm}\psfig{figure=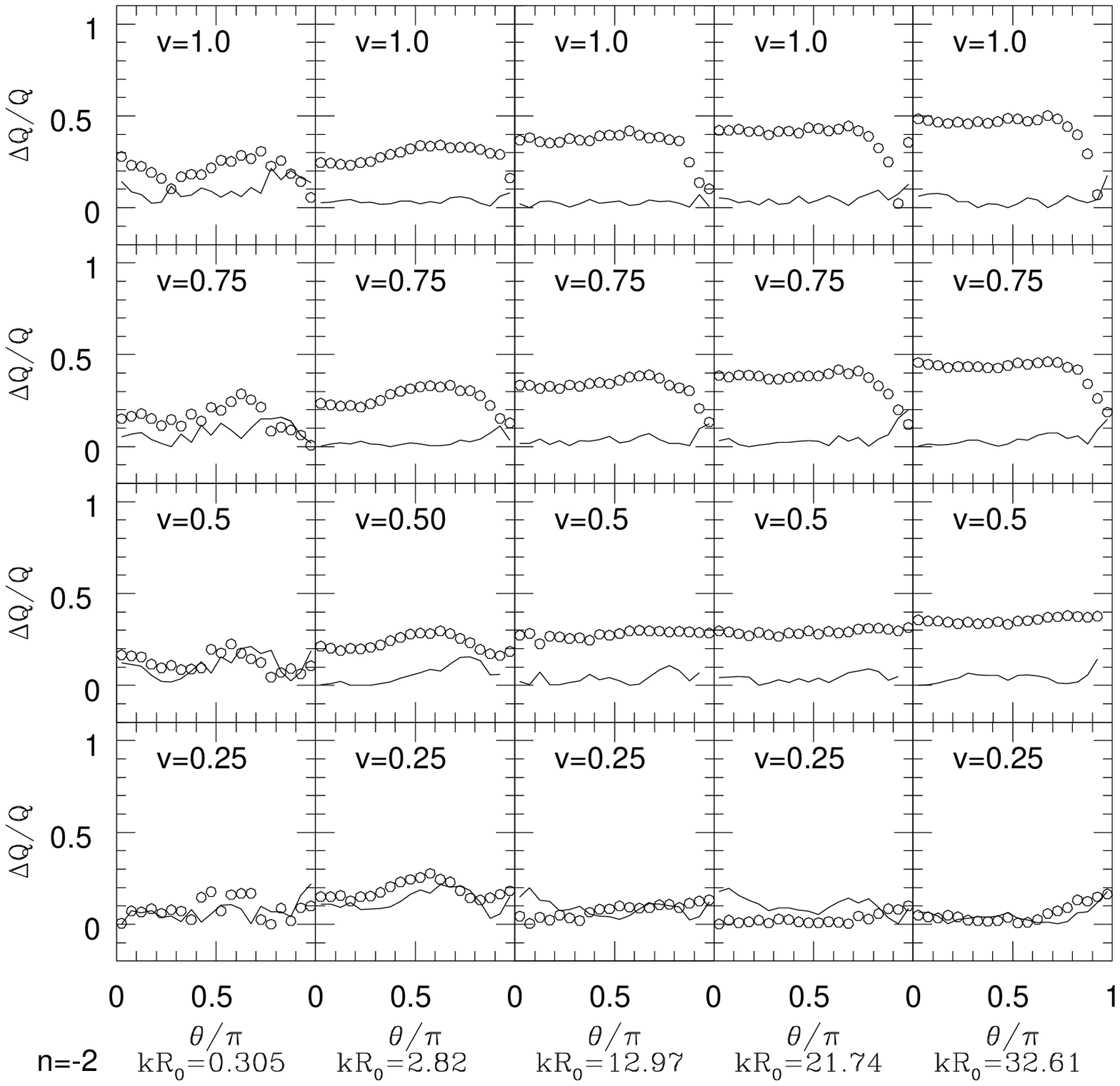,width=14.cm}}
\caption{Same as Fig.\ref{erro2910}, but for the spectral index
$n=-2$.}
\label{erro2930}
\end{figure}

To quantify the accuracy of our fitting formula, we plot the
percentage of the deviation $\Delta Q/Q$ in Figures~\ref{erro2910},
\ref{erro2920} and \ref{erro2930}.  The deviation is defined as:
\begin{equation}
{\Delta Q \over Q}={\mid Q_{fit}-Q_{simu}\mid \over Q_{simu}},
\end{equation}
where $Q_{fit}$ is the prediction of our fitting formula, and
$Q_{simu}$ is the reduced bispectrum measured from the
simulations. Similarly we also estimate the accuracy of the fitting
formula of SF99 that is also shown in the figures. The figures show
that our fitting formula can match the measured results typically at
an accuracy of 10 percent. The deviation is slightly larger, about
$20\sim 30\%$, when $\theta\approx \pi$ and $k R_0< 1$, which could be
attributed to some stochastic fluctuation in $Q_{simu}$ for the
limited number of k-triangles in the configuration.  We also see that
the deviation for the SF99 formula is much larger, almost about 50
percent in the strongly nonlinear regime.

Although we only have a single realization for these simulations, we
measured the bispectrum for two or three outputs scaled according to
the similarity solution. We regard each output as a realization, and
estimate the errors of the bispectra from the different outputs. We
also have taken the projections along different axes as independent
realizations when we estimate errors for the 2D bispectra.  In
Figures~\ref{fitting2910}, \ref{fitting2920} and \ref{fitting2930},
the error bars are estimated with this method.
 
\section{Summary and discussion}
In this paper, we have accurately measured the bispectrum for four
scale-free models. The measurement is based on a new method that can
effectively eliminate the alias and numerical artifacts, and reliably
extend the analysis into the strongly non-linear regime. The work also
makes use of a set of state-of-the art N-body simulations of
scale-free hierarchical models that have a significantly larger
dynamical range than the previous studies. With these measured
results, we demonstrated that the measured bispectrum depends on the
shape and size of $k$-triangle even in the strongly nonlinear
regime. It increases with wavenumber and decreases with the spectral
index.  These results are consistent with that those found for the
three-point correlation for CDM models (Jing \& B\"orner 1998, Ma \&
Fry 2000, Scoccimarro \& Couchman 2001 , Scoccimarro et al. 2001b,
 Takada \& Jain 2003a), but are in
contrast with the hypothesis that $Q$ is a constant in the
strongly non-linear regime (SF99). We also show that the fitting
formula of SF99 does not describe our simulation results well, with a
typical error about 40 percent.  In the end, we present a new fitting
formula for the reduced bispectrum that is valid for $-2 \leq n \leq 0$
 with a typical error of $10$ percent only.

Our new method for measuring the bispectra is to use the property that
the 2D power spectrum and bispectrum are identical to the 3D
ones. This property can be easily proved and has been tested with our
N-body simulations. As the 2D FFT requires less computer memory than
the 3D FFT, we can extend our analysis of the bispectrum into very
small scales with little computer limitation.  We also use the scaling
properties of the scale-free models to correct for all known numerical
artifacts and to identify those regimes where the bispectra can
reliably measured.

Although we have obtained an empirical formula of the bispectrum for
scale-free models of the spectral index $-2 \leq n \leq 0 $, some
issues need to be addressed in future work. One obvious aspect is that
the formula does not work for $n=1$, because the tree-level PT cannot
predict the weakly non-linear bispectrum for this model. Fortunately
the slope of the power spectrum in CDM models, which are the most
plausible theory for the structure formation in the Universe, is in
the range $-1$ to $-3$ at the non-linear scales. Therefore, we may
generalize our fitting formula to CDM models by considering the
change of the power spectrum slope with scale as well as the
deviation from the Einstein-de Sitter model. We will study the
bispectrum in CDM models in a future paper.  Another issue is that our
fitting formula is purely empirical. Considering that our measured $Q$
results are in good agreement with the one-loop PT prediction in the
quasilinear regime (Scoccimarro et al. 1998) and that the halo model
can successfully match the three-point correlation function in the
strongly non-linear regime (Takada \& Jain 2003a), we think it would
be possible to combine these theoretical predictions to find a more
theory-oriented fitting formula for $Q$.

\acknowledgments
We thank Ue-Li Pen for useful discussion at the initial stage of this work.
The work is supported in part by NKBRSF (G19990754), by NSFC
(Nos.10125314, 10373012), and by the CAS-MPG exchange
program. Numerical simulations presented in this paper were carried
out at ADAC (the Astronomical Data Analysis Center) of the National
Astronomical Observatory, Japan.

\clearpage

\begin{table}
\begin{center}
\begin{tabular}{cccccccc}
\hline\hline
 simulation & $\eta /L$ & timesteps& $a_{initial}$& $a_{first output}$ &$a_{final}$&outputs&$\Delta\log a_i$\\
\hline
n=1.  &$1.\times 10^{-4} $&2000&0.0007&0.0042&1.0 & 10 &0.266\\
\hline
n=0.  &$1.\times 10^{-4} $&2000&0.0028&0.0157&1.0 & 10 &0.20\\
\hline
n=$-1$&$1. \times 10^{-4}$&2000&0.0136&0.064 &1.0 & 11 &0.121\\
\hline
n=$-2$&$1. \times 10^{-4}$&2000&0.0834&0.2514&1.0 & 10 &0.0667\\
\hline\hline
\end{tabular}
\end{center}
\caption{The scale-free simulations of $512^3$ particles}
\end{table}

\end{document}